\documentclass[doublecol]{epl2}
\usepackage{epsfig,color,multirow}
\begin{document}

\title{Violation of Onsager symmetry for a ballistic channel Coulomb coupled to a quantum ring}

\author{B. Szafran\inst{1} \and M.R. Poniedzia{\l}ek\inst{1} \and F.M. Peeters{\inst2}}
\shortauthor{B. Szafran, M.R.Poniedzia{\l}ek, \and F.M. Peeters}
\institute{Faculty of Physics and Applied Computer Science, AGH
University of Science and Technology,\\ al. Mickiewicza 30, 30-059
Krak\'ow, Poland}
\institute{
\inst{1}Faculty of Physics and Applied Computer Science, AGH University of Science and Technology, al. Mickiewicza 30, 30-059
Krak\'ow, Poland  \\
\inst{2}
Departement Fysica,  Universiteit Antwerpen,
Groenenborgerlaan 171, B-2020 Antwerpen, Belgium
}

\abstract{ We investigate a scattering of electron which is injected
individually into an empty ballistic channel containing a cavity
that is Coulomb coupled to a quantum ring charged with a
single-electron.
 We solve the time-dependent Schr\"odinger equation for
the electron pair with an exact account for the electron-electron
correlation. Absorption of energy and angular momentum by the
quantum ring is not an even function of the external magnetic field.
As a consequence we find that the electron backscattering
probability is asymmetric in the magnetic field and thus violates
Onsager symmetry. } 
 \maketitle

\section{Introduction}

The transport properties of two-dimensional mesoscopic and
nano-scale conductors are usually studied in external magnetic field
$B$ applied perpendicular to the plane of electron confinement. The
current / voltage ($I/V$) characteristics of the device is described
by its conductance ($G$), $I(B,V)=G(V,B)V$. In the linear transport
regime the conductance is independent of $V$,  moreover it is an even
function of the magnetic field \cite{b} $G(B)=G(-B)$  -- relation
that is known as the Onsager symmetry.

In the Landauer approach the linear conductance $G(B)$ is determined by the
probability  $T(B)$ that a Fermi-level electron is transferred from one
terminal to the other $G(B)={e^2}T(B)/h$.
 The  dependence of the transfer probability on $B$ results from Aharonov-Bohm
 interference and from deflection of the electron trajectories by the Lorentz force.
Due to the magnetic forces the kinetics of the electron transfer
through an asymmetric channel is different for opposite magnetic
field orientations. However, the backscattered trajectories are
identical for $\pm B$ which results in the Onsager symmetry.

In the non-linear transport regime the magnetic-field symmetry of
the current can be broken \cite{sb}. The magnetic asymmetry of the
non-linear current was observed in various systems, including open
quantum rings \cite{qr} and dots (cavities) \cite{qd} as well as in
carbon nanotubes \cite{cn}. A number of scenarios for the appearance
of the asymmetric current were given, including potential landscape
being not an even function of the magnetic field \cite{sb,poli},
effects of the electron-electron interaction within the channel
\cite{ci} or capacitive coupling of the channel to the other
conductor which is driven out of equilibrium \cite{sk} by an applied
bias.


The studies of the magnetic field asymmetry of the current
\cite{sb,qr,qd,cn,poli,ci,sk} concerned standard devices which are
filled by the electron gas.  Recently, Gustavsson et al. \cite{last}
studied experimentally the Aharonov-Bohm self-interference of electrons injected into the
channel one by one. In the experiment single-electron injection
occurs due to Coulomb blockade which allows only a single electron
to move between the source and the drain. The single-electron valve
applied in the device \cite{last} is a quantum dot occupied by a
single-electron with the energy below the Fermi energy of the
source. The electron can be ejected from a bound quantum dot state
into the channel only using the excess energy of another electron
which tunnels from the source Fermi level to replace the ejected
electron in the quantum-dot-confined state (so called co-tunneling
process). In the experiment \cite{last} the transmission of
electrons through the device is detected by a counter instead of
ammeter applied in standard experiments that measure the current
carried by the electron gas. The electron counter consists of a
quantum point contact charge coupled to the channel. In the
single-electron injection regime the counter registers transmission
of only about a hundred electrons per second, which is due to the
low probability of the co-tunneling process.

The separation of a single electron of the two-dimensional electron
gas was accomplished already in the nineties but only in the quantum
dot confined state \cite{ash}. With the single-electron
injection\cite{last} it becomes possible to study a single electron
in motion. The experiment with single electrons moving across an empty
channel provides an attractive alternative to standard experiments
where the current is carried by a gas of electrons, or in the linear
regime, by electrons at the Fermi level. The single-electron
scattering conditions resemble the ones present in atomic physics
but in the solid-state scattering experiment the potential may be
formed at will and a single scattering center is naturally observed.
  Since the moving electron is the only one inside
the channel it is no longer shielded from the environment like in
the case of a two-dimensional electron gas. The moving electron will
be therefore more effectively coupled to any external charges. In
the linear transport regime electrons at 0 K temperature are not
able to loose energy because of the exclusion principle. This is
different for a single electron that moves through an empty channel
which, because of the interaction with the environment, can loose
energy and momentum and thus undergo inelastic scattering.

In the present paper we consider an electron moving in an empty
semiconductor nanochannel and its scattering by a potential cavity
that is capacitively coupled to a quantum ring containing a single
electron. The electron is injected into the channel i) from a bound localized
electron state formed within a quantum dot ii) with a finite kinetic
energy (that in the experiment \cite{last} is equal to the
difference between the Fermi level of the source and quantum dot
energy level) and iii) the electron moves across an empty channel,
i.e. not filled by an electron gas, iv) the injection due to the
release of the electron from the bound-state has intrinsically a
time-dependent character in contrary to the stationary flow of the
current at the Fermi level. We develop a scheme to simulate the single-electron scattering based
on the time-dependent Schr\"odinger equation. Calculations are
numerically ''exact" with full account of the correlation between
the electron within the channel and the electron confined in the
quantum ring. We find that the electron transmission probability is
not symmetric in $B$. The asymmetry is due to the inelastic
scattering properties of the quantum ring which tends to absorb the
angular momentum of the channel electron. We therefore indicate a
new mechanism for violation of the Onsager symmetry specific for the
single-electron injection conditions.

\begin{figure}[ht!]
\epsfysize=70mm \epsfbox[28 100 518 612]{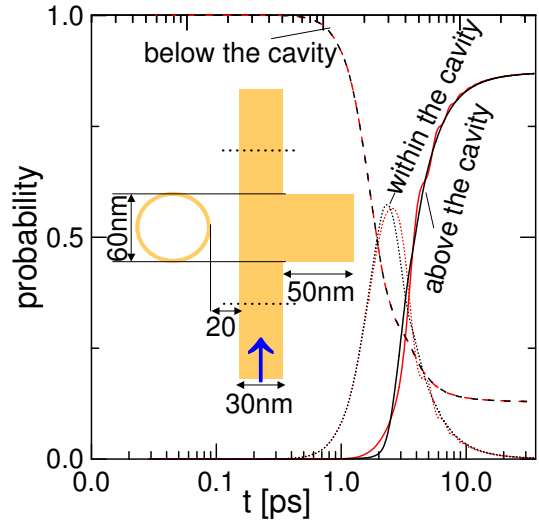} \\
\caption{ The inset shows the schematic drawing of the system. The
electron is injected from below. Magnetic forces tend to deflect the
electron trajectory to the left (right) for $B>0$ ($B<0$). The main
panel shows the probability to find an electron inside (dotted
curves), above (solid curves) and below (dashed curves) the cavity
for $B=0.7$ T (red curves) and $B=-0.7$ T (black curves) as function
of time obtained for uncharged quantum ring. The cavity region is
the one between thick dotted lines in the inset. } \label{1sc}
\end{figure}

\section{Theory}

\revision {Although the present paper is inspired by the 
single-electron injection of Ref. \cite{last} we consider here a different geometry.
The ring is formed beside and not within the channel. }
The studied system is schematically depicted in the inset of Fig. 1.
The width of the channel is locally increased near the ring.  The channel edges
are taken as infinite potential barriers.
Spacing between the center of the ring and the left edge of the
channel is equal to 50 nm. Deformation of the electron density
within the ring has an angular character, hence the ring is assumed
strictly one-dimensional. 

We consider the two-electron Hamiltonian
\begin{equation} H=h_1+h_2+V_{ee}(|{\bf
r}_1-{\bf r}_2|),\end{equation} where \begin{equation} h_n=\left(-i\hbar\nabla_n+e{\bf A}({\bf
r}_n)\right)^2/{2m^*}+F({\bf r}_n)\end{equation} is the single-electron energy
operator ($n=1,2$) and $F$ is the potential defining the channel.
The electron-electron interaction is assumed to be a screened
Coulomb interaction
$V_{ee}(r)=\frac{e^2}{4\pi r\epsilon\epsilon_0}\exp(-r/\lambda)$ with
the screening length
$\lambda=500$ nm, which is large in comparison to the cavity \& ring
system. GaAs dielectric constant $\epsilon=12.9$ and
electron effective band mass $m^*=0.063m_0$ are applied.

 We neglect the tunneling between the ring and the
cavity. Therefore, the two electrons that occupy separate regions in
space -- channel and the ring -- can be treated as distinguishable.
The single-electron Hamiltonian eigenstates of the electron in the ring $\phi_l({\bf r}_1) $ and
the electron inside the channel $f_k({\bf r}_2)$ form a complete basis for the two-electron problem. The solution
of the time-dependent Schr\"odinger
equation $i\hbar d\Psi/dt=H\Psi$ can be therefore written in a general form\cite{uwaga}
\begin{equation}\Psi({\bf r}_1, {\bf r}_2,t)=\sum_{kl} c_{kl}(t) \phi_l ({\bf r}_1) f_k({\bf r}_2),\end{equation}
in which the entire time dependence is contained in the set of coefficients $c_{kl}$.
We define a partial wave packet of the electron within the  channel associated with the quantum ring state of angular momentum $l$
\begin{equation}
\psi_l({\bf r_2},t)=\sum_{k} c_{lk}(t)\phi_k({\bf r}_2).
\end{equation}
With this definition the general solution of the Shr\"odinger equation acquires the form
\begin{equation}\Psi({\bf r}_1, {\bf r}_2,t)=\sum_l \phi_l ({\bf r}_1) \psi_l ({\bf r}_2,t).\end{equation}
The wave functions of the electron inside the ring  written in the
symmetric gauge $A=\frac{B}{2}\left(y_c-y,x-x_c,0\right)$ [the ring
center coordinates are $(x_c,y_c)$] have the form
$\phi_l=\frac{1}{\sqrt{2\pi}}\exp(il\theta)$. Equations for the time
evolution of the partial wave packets are obtained by inserting Eq.
(5) into the Schr\"odinger equation followed by a projection on a
quantum ring state of angular momentum $k$
\begin{eqnarray}
 \frac{i\hbar\partial \psi_k({\bf r}_2,t)}{\partial t}=\sum_l \left[\left(E_l+h_2\right)\delta_{lk}+ W_{kl}({\bf r_2})\right]\psi_l({\bf r}_2,t).\end{eqnarray}
In the above equation
$E_l=\frac{\hbar^2}{2m^*}\left(l+\Phi/\Phi_0\right)^2$ is the energy
level of the isolated quantum ring [see Fig. 2(b)] with $\Phi_0=h/e$
-- the magnetic flux quantum and $\Phi=B\pi R^2$.  In Eq. (6) the
interaction term \begin{equation} W_{kl}({\bf
r_2})=\langle\phi_{k}({\bf r}_1)|V_{ee}(|{\bf r}_1-{\bf
r}_2|)|\phi_{l}({\bf r}_1)\rangle.\end{equation} induces
oscillations between partial wave packets corresponding to different
angular momenta of the electron within the ring. We solve a system
of five equations (6) taking into account $l=0,\pm1,\pm2$ states of
the ring. For magnetic fields studied below contribution of the
partial wave packets with higher angular momenta is negligible. The
system of equations (6) is solved using the Crank-Nicolson scheme
with a finite difference approach in which we apply the
gauge-invariant discretization of the kinetic energy operator
introduced in Ref. \cite{Gov}. The entire computational channel is
taken $20$ $\mu$m long with the mesh spacing of $2$ nm.
 The initial state for the electron within the channel $\chi({\bf r}_2)$
is prepared in the way previously applied in Ref. \cite{naszeprl}:
within the channel we form a shallow rectangular potential well of
depth 5 meV with the center 200 nm below the cavity of Fig. 1 and a
spread along the channel of $\pm 80$ nm. The electron in the quantum
well is relaxed to the ground-state. Then the potential well is
removed and the packet is pushed up along the channel by a pulse of
an electric field acting till the electron acquires an average
momentum $\langle-i \hbar \frac{\partial}{\partial y} +e
A_y\rangle=\hbar k$, for $k=0.04$/nm. This value corresponds to a
kinetic energy of progressive motion $\hbar k^2/2m^*=0.91$ meV.

We assume in the initial condition that the electron
in the ring is in the ground-state of $l$ angular momentum before the scattering process,
hence we apply  $\psi_l=\chi$ and $\psi_k=0$ for $k\neq l$.
\revision{Admixtures of the excited ring states are present in the final state. 
The scattering process that we describe in this letter lasts about 50 ps.
In the considered energy range (sub meV) the dominant relaxation mechanism
is due to the electron coupling to acoustic phonons. The relaxation times
of the excited ring states to the ground state are of the order of nano seconds \cite{Piacente},
which allows us to neglect energy and angular momentum relaxation during
the scattering. 
On the other hand, the transmission rates in the experiment of Gustavsson et al. \cite{last}
vary between 10 Hz (single-electron injection regime) and 10 kHz (sequential tunneling regime).
The electron inside the ring will therefore relax to the ground-state before the next electron
enters the channel, which supports the initial condition applied in the present calculation.}

\section{Results}

Probability density for the electron within the channel is
calculated as $\sum_l |\psi_l({\bf r_2},t)|^2$. The time-dependence
of the part of the packet before, within, and above the cavity for
neglected electron-electron interaction is plotted in Fig. 1 for
$B=\pm 0.7$ T. The part below the cavity -- incoming and reflected
parts of the packet -- is identical for both magnetic field
orientations at any moment in time which is a result  of the
invariance of the backscattered trajectories with respect to the $B$
vector orientation. The parts of the packet within the cavity and
above it for opposite $B$ orientations are not identical due to the
Lorentz force effect \cite{time}. At the end of the simulation the
cavity is left empty and as a consequence in the large $t$ limit the
same transfer probability is obtained for $\pm B$ in large $t$
limit, leading to the Onsager symmetry for the transfer probability
$T(B)=T(-B)$ \cite{time}.

\begin{figure}[ht!]
\epsfysize=70mm  \epsfbox[28 80 518 612]{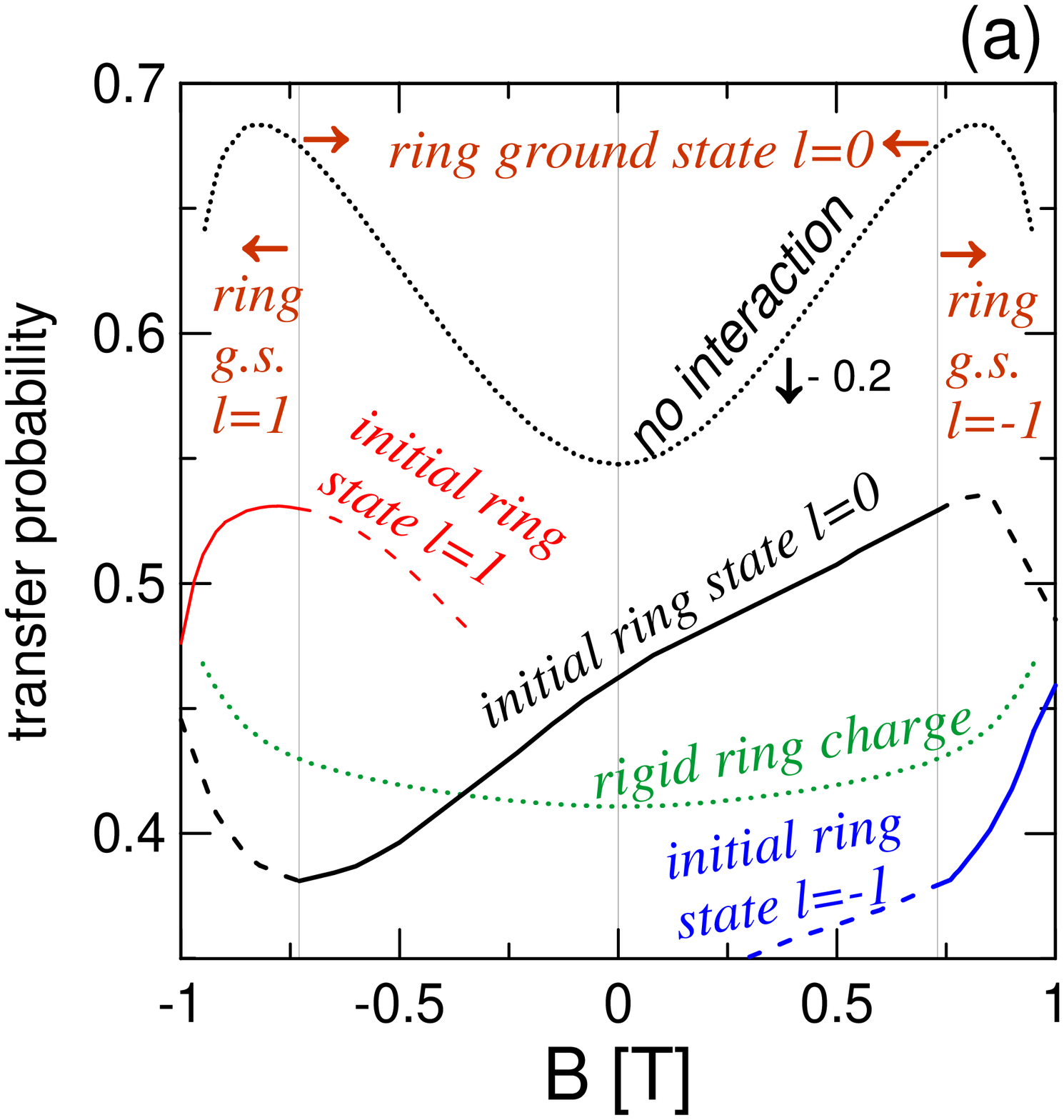} (a) \\
\epsfysize=45mm  \epsfbox[33 298 576 605]{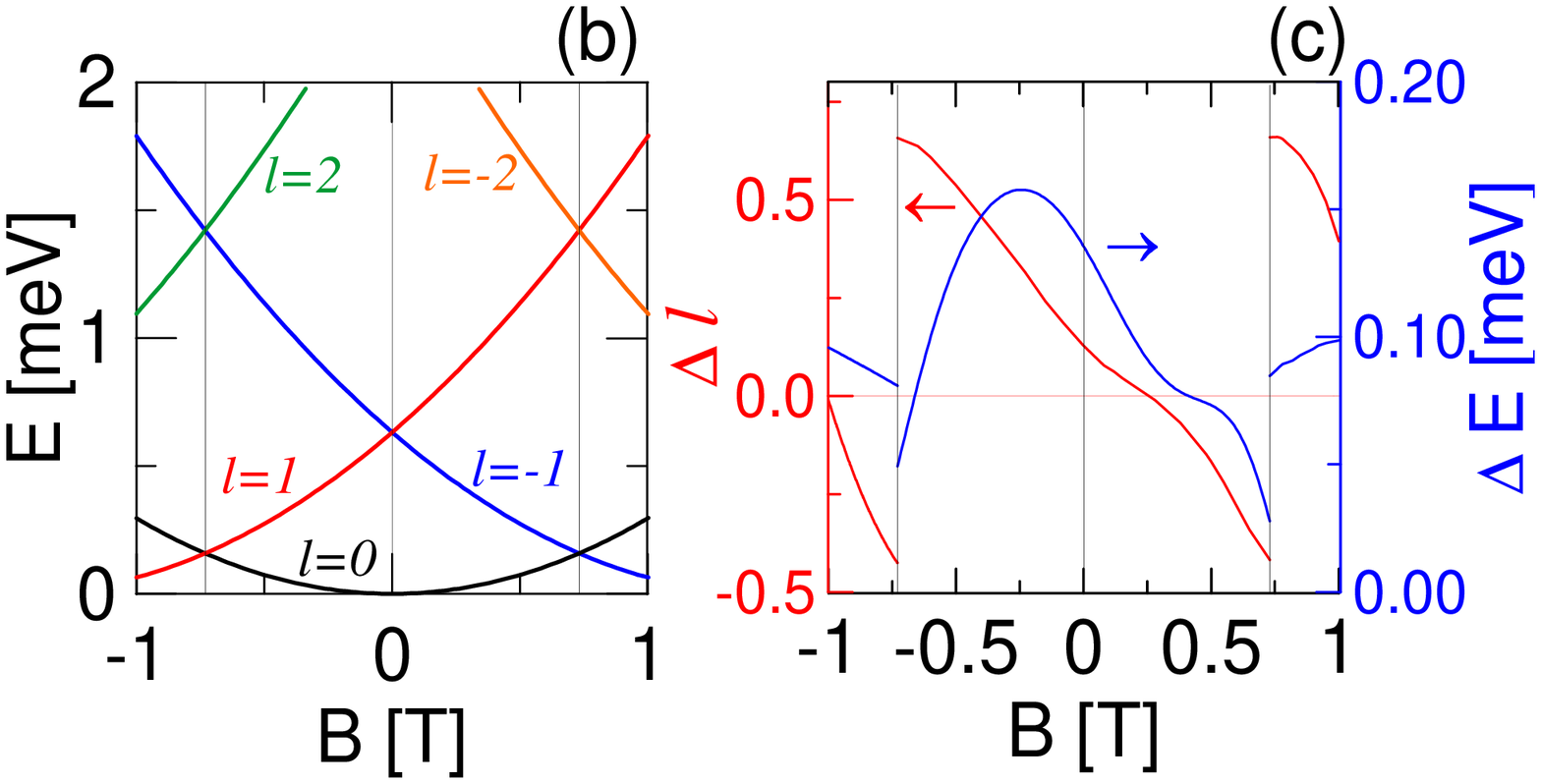} \\
\epsfysize=50mm  \epsfbox[28 66 561 625]{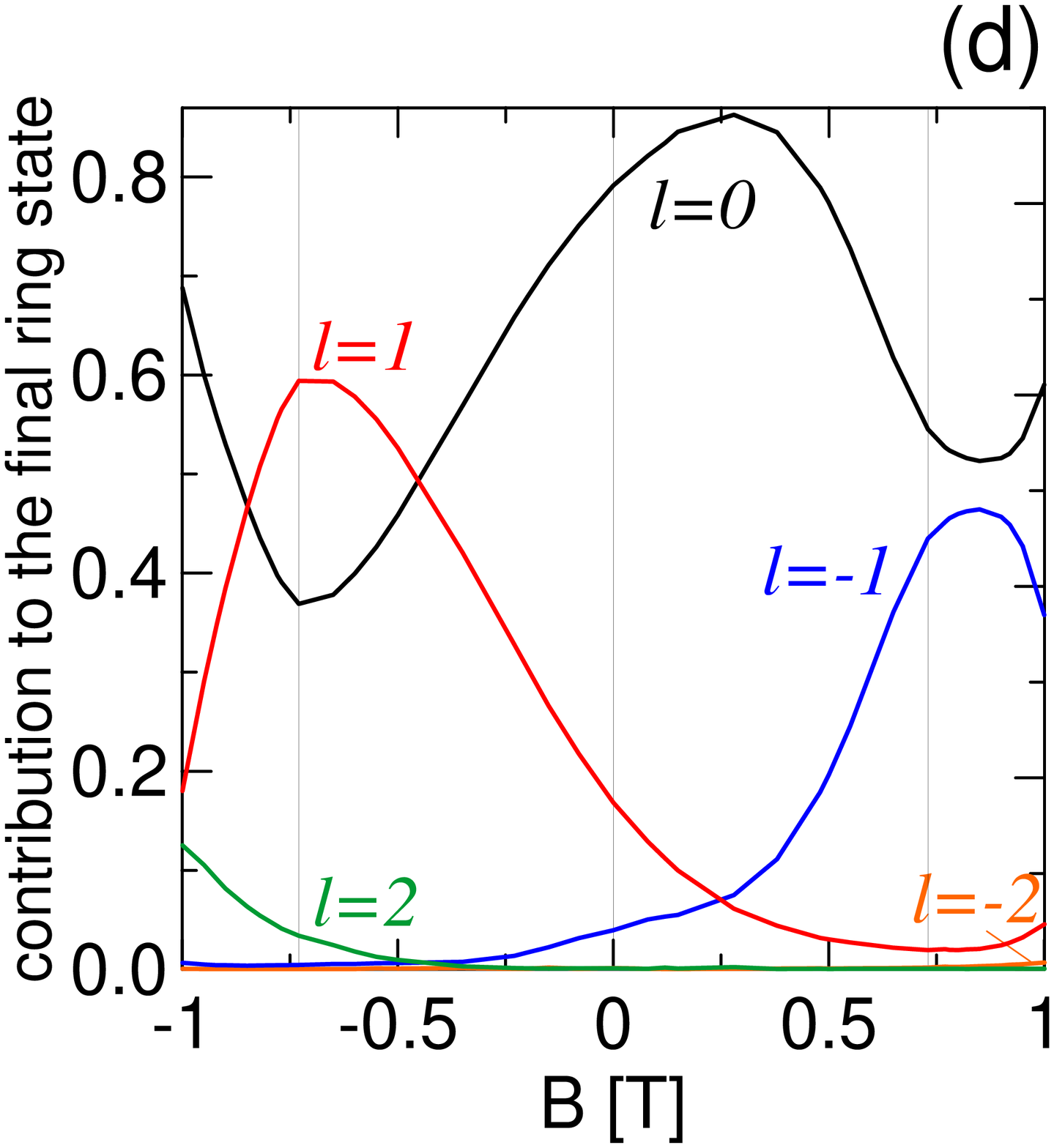} \\
\caption{(a) Electron transfer probability for an empty quantum ring
(black dotted curve - shifted down by 0.2) and for rigid ring charge
(green dotted curve). Black, red and blue curves are for the initial ring state $l=0$, $l=1$ and $l=-1$,
respectively. Solid (dashed) parts of these curve correspond to $B$
in which $l$ is the ground (excited) quantum ring state.
(b)
Energy spectrum of an isolated ring. (c) Angular momentum (red curve) and energy
(blue curve) absorbed by the quantum ring (difference of final and
initial values) for the ring ground-state used as initial condition.
(d) Contributions to the final ring state for the initial ring state
$l=0$. In all the plots the vertical lines indicate $B=0$
as well as magnetic fields for which
the ground-state angular momentum transition occurs in the ring $B=\pm 0.73$ T.
} \label{t}
\end{figure}

$T(B)$ (shifted down by -0.2) for neglected interaction between the
ring and the channel is plotted in Fig. \ref{t}(a) by the black
dotted curve. The blue dotted curve in Fig. \ref{t} shows the
results when the interaction is included but for a rigidly uniform
density of the electron within the ring (the result corresponds to a
single equation (1) for the basis limited to $l=0$).
 The electron-electron repulsion reduces the transfer probability
but $T$ -- for the frozen quantum ring state - remains an even function of $B$.

For the frozen quantum ring state the channel electron is scattered elastically.
The inelastic scattering and the reaction of the ring electron to the
Coulomb repulsion of the channel electron
is taken into account for the complete basis of $l$ states included in the basis.
The black curve in Fig. \ref{t} shows the results
 for the complete basis obtained for the $l=0$ initial state of the quantum ring
[this is the isolated ring ground-state for
$\Phi\in(-\Phi_0/2,\Phi_0/2)$ or $B\in (-0.73,0.73)$ T - see Fig.
2(b)].  At $B=0$ the transfer probability is increased with respect
to the rigid ring charge case. Near zero magnetic field  $T(B)$
distinctly grows with $B$ and the Onsager symmetry is
violated. 

\begin{figure*}[ht!]
\epsfxsize=150mm  \epsfbox[54 329 515 751]{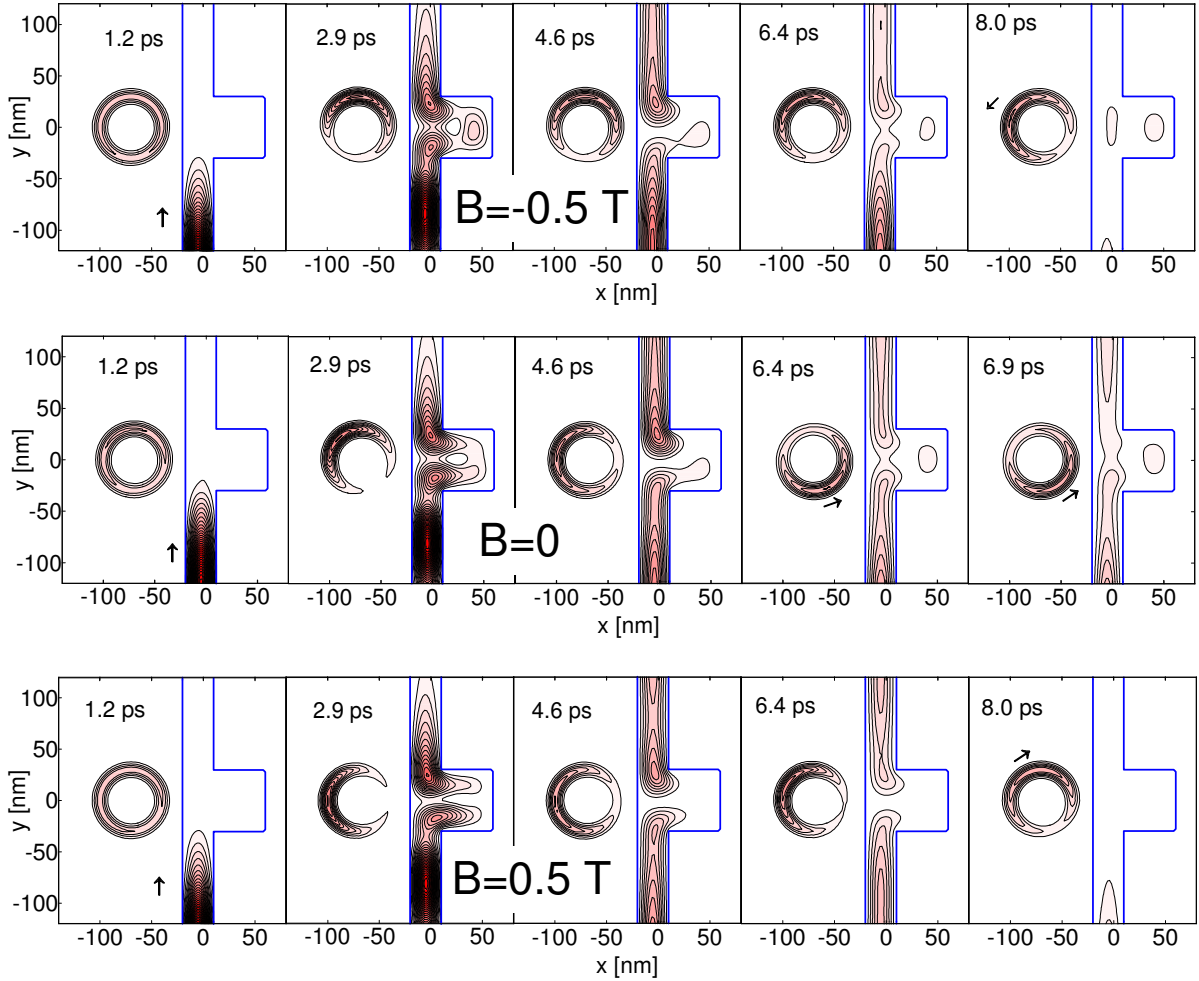} \\
\caption{Snapshots of the electron density within the channel and
inside the ring for $B=0$ and $\pm 0.5$ T for several moments in time. The
boundaries of the channel are marked with the blue line. The quantum
ring electron density is first deformed (2.9 ps). The resulting state
is a non-stationary mixture of $l$ eigenstates that
rotate counterclockwise for $B=-0.5$ T and for $B=0$ (positive average angular momentum with respect to
the center of the ring) and clockwise for $B=0.5$ T . Color scales for the ring and channel
electron densities are not the same. } \label{rys}
\end{figure*}

Snapshots of the evolution of the charge density for $B=0$ and
$B=\pm 0.5$ T is plotted in Fig. \ref{rys}. The electron density
within the ring is first strongly deformed by the repulsion of the
incoming electron.
After the scattering the electron state within the ring is a mixture
of $l$ eigenstates and its charge density rotates counterclockwise
in case of  $B=-0.5$ T and $B=0$ and clockwise for $B=0.5$ T in
accordance with the sign of the average value of the angular momentum as
presented in Fig. 2(c).

Fig. 2(d) shows the contributions of different angular momentum ring
states to the final state for the case that the quantum ring
electron occupied initially the $l=0$ orbital as obtained when the channel electron has left the interaction
region. For $B=0$ we notice that although the $l=+1$ and $l=-1$
quantum ring states are degenerate [Fig. 2(b)] in the final state
the contribution of angular momentum $l=+1$ dominates over the one
for $l=-1$ -- see also for the difference of final and initial ring
angular momentum depicted in Fig. 2(c). The incoming electron
carries positive angular momentum $(\simeq 2\hbar)$ with respect to
the center of the ring. Its partial backscattering is related to the
loss of angular momentum that it carries. For the studied system the
total angular momentum is not conserved, however the quantum ring at
$B=0$ tends to absorb the angular momentum of the channel electron.
For $B>0$ the persistent current of $l=1$ state produces a magnetic
dipole moment which is oriented antiparallel to the external field
\cite{physicae} hence $E_1$ grows with $B$ [Fig. 2(b)]. As a
consequence an increase of $B$ is accompanied by an increase of the
energy cost of quantum ring excitation from $l=0$ to $l=1$ which is
responsible for the partial absorption of angular momentum. The
transfer of the angular momentum to the ring is reduced [Fig. 2(c)],
and so is the backscattering probability [Fig. 2(a)]. Consistently,
for $B<0$ the transfer of angular momentum to the quantum ring and
the backscattering  are enhanced.

The angular momentum absorbed by the ring [Fig. 2(c)] is more or
less a linear function of $B$ within the range of $l=0$ quantum ring ground-state.
However, the absorbed energy is not a
monotonic function of $B$. For $B<0$ the energy absorbed by the
quantum ring first increases due to the increasing contribution of
the excited $l=1$ state to the final ring state. For $B<-0.25$ T the
energy absorbed by the ring starts to decrease due to lowered energy
cost of the excitation from $l=0$ to $l=1$  [Fig. 2(b)].

For $B>0.73$ T ($B<-0.73$ T) the ring ground state corresponds to
$l=-1$ $(l=1)$. For the quantum ring electron assumed in the
ground-state before the scattering  $T(B)$ possesses discontinuities
at $B=\pm 0.73$ T -- see the solid parts of the red, black and blue
curves in Fig. 2(a). For $B>0.73$ T the lowest-energy excitation of
the ring-confined electron is related to the transition
$l=-1\rightarrow 0$, i.e. to the positive change of the angular
momentum carried by the ring. The quantum ring can again absorb the
angular momentum  of the channel electron (see the jump in the red
line in Fig. 2(c)) which is related to  the rapid drop of $T$ above
the quantum ring ground-state transformation (see the solid black
line below 0.73 T and the solid blue line above 0.73 T). Opposite
effect is observed at the transition near $B=-0.73$ T. For $B<-0.73$
T the quantum ring ground-state is transformed to $l=1$. In order to
absorb the angular momentum of the channel electron the ring
electron should be excited to $l=2$ state which is still high in
energy [see Fig. 2(b)], hence the jump of $T$ in Fig. 2(a) when $B$
decreases below $-0.73$ T (solid red line at left of the critical
field and solid black line at its right).

\section{Discussion}

Magnetic field asymmetry of the single-electron transmission through
an asymmetric channel was previously considered in Ref. \cite{kali}.
It was demonstrated that the transmission probability becomes an
asymmetric function of the external magnetic field when the channel
is coupled to a metal plate placed at a distance above the channel.
The coupling of the electron in the channel with the electron gas in
the metal is due to the induction of a positive charge on the metal
surface that acts on the electron like the image charge. The image
potential translates the different kinetics of the wave packet
transfer for opposite magnetic field orientations into a different
effective scattering potential. The obtained $T(B)$ asymmetry was
attributed \cite{kali} to the nonlinear character of the
Schr\"odinger equation with the image charge potential. The
non-linear electron-metal interaction -- similar to the Hartree
potential --  results from the assumption that one can attribute a
separate wave function to the electron within the channel, which
implies a complete neglect of the quantum correlations with the
electrons in metal. In the present work we considered an electron
within the channel that was charge coupled to a quantum ring
containing a single-electron. For that system the problem can be
treated exactly and we developed a scheme to solve the linear
Schr\"odinger equation with full account taken of the correlation
between the electron within the channel and the electron the ring. A
distinct violation of the transmission probability symmetry found
here occurs already near $B=0$ and results in asymmetric inelastic
scattering properties of the ring, while the asymmetry of Ref.
\cite{kali} -- due to the image charge reinforcement of the Lorentz
force effects -- was visible only at relatively high fields. The
Lorentz force, although included in the present model, has a weak
effect on the transmission process (Fig. 1) due to the relatively
small size of the cavity. Moreover, in the applied coplanar geometry
the electron inside the ring can reinforce the Lorentz force effects
only to a limited extent, much smaller than the metal placed above
the structure as in Ref. \cite{kali}.

The potential acting on the electron within the channel as described
in the present approach is linear although non-scalar. Therefore,
the non-linearity of the Schr\"odinger equation is not a necessary
condition for the magnetic field asymmetry of the single-electron
scattering process. The asymmetry results from the interaction of
the channel electron with a non-rigid environment.

\section{Summary and conclusions}
In summary, we have studied the scattering of a single-electron injected into
a channel containing a potential cavity Coulomb coupled to a charged quantum ring.
 The problem
was solved with the time-dependent Schr\"odinger approach. The
proposed method naturally accounts for inelastic scattering of the channel electron.
The obtained results
indicate that the backscattering of the electron is enhanced
whenever the quantum ring can absorb the angular momentum of the
channel electron.  Due to the form of the single-ring spectrum the
absorption of angular momentum is enhanced for one $B$ orientation
and reduced for the other, hence leading to the violation of the Onsager symmetry
for a single electron scattering event.

\section{Acknowledgements} This work was supported (in part) from the
AGH UST project 11.11.220.01 "Basic and applied research in nuclear
and solid state physics", the Belgian Science Policy (IAP) and the
Flemish Science Foundation (FWO-VI). Calculations were performed in
ACK\---CY\-F\-RO\-NET\---AGH on the RackServer Zeus.


\begin{thebibliography}{00}
\bibitem{uwaga} The antisymmetrization of $\Psi$ for electrons occupying separate regions
in space is straightforward to perform, but we checked that it does not affect any of the found physical properties
of the system. Therefore we skipped it.
\end{thebibliography}
\end{document}